\documentclass[3p,times,procedia]{elsarticle}
\usepackage{nupha_ecrc}

\volume{00}

\firstpage{1}

\journalname{Nuclear Physics A}

\runauth{}

\jid{nupha}

\jnltitlelogo{Nuclear Physics A}

\usepackage{amssymb}

\usepackage[figuresright]{rotating}

\begin{document}

\begin{frontmatter}



\dochead{XXVIIth International Conference on Ultrarelativistic Nucleus-Nucleus Collisions\\ (Quark Matter 2018)}

\title{Probing the transverse size of initial inhomogeneities with flow observables }


\author[label1]{Fernando G. Gardim}
\author[label2]{Fr\'ed\'erique Grassi}
\author[label2]{Pedro Ishida}
\author[label2]{Matthew Luzum}
\author[label1]{Pablo S. Magalh\~aes}
\author[label3]{Jacquelyn Noronha-Hostler}
\address[label1]{Instituto de Ci\^encia e Tecnologia, Universidade Federal de Alfenas, 37715-400 Po\c{c}os de Caldas-MG, Brazil}
\address[label2]{Instituto de F\'{i}sica, Universidade de S\~ao Paulo, Rua do Mat\~ao 1371,  05508-090 S\~ao Paulo-SP, Brazil}
\address[label3]{Department of Physics and Astronomy,
Rutgers, The State University of New Jersey, Piscataway, NJ 08854-8019, USA}

\begin{abstract}
 Disentangling the effect of initial conditions and medium properties is an open question in the field of relativistic heavy-ion collisions. We argue that, while one can study the impact of initial inhomogeneities by varying their size, it is important to maintain the global properties fixed.
 We present a method to do this. We show that many observables are insensitive to the the hot spot sizes, including integrated $v_n$, scaled distributions of $v_n$, symmetric cumulants, event-plane correlations, and differential $v_n(p_T)$. We find however that the factorization breaking ratio $r_n$ and sub-leading component in a Principal Component Analysis are more sensitive to the initial granularity and can be used to probe short-scale features of the initial density.

\end{abstract}


\end{frontmatter}


\section{Objective}
Relativistic viscous hydrodynamic models have been successful in describing dynamics of the Quark Gluon Plasma within in the context of heavy-ion collisions.  For these models, initial conditions that describe the brief far-from-equilibrium state immediately after the collisions are needed and heavily influence the final flow harmonics measured in experiments.  Depending on the underlying physical assumptions of these initial conditions, both the geometrical large scale structure quantified through eccentricities and the small scale structure known as ``hot spots" may differ. In this work, we study the influence that the size of these hot spots have on measured observables while maintaining the eccentricities approximately constant, further details can be found in \cite{Gardim:2017ruc}.

\section{Method: smoothing of initial conditions}

\begin{figure}[!ht]
  \begin{center}
\includegraphics[width=0.6\textwidth]{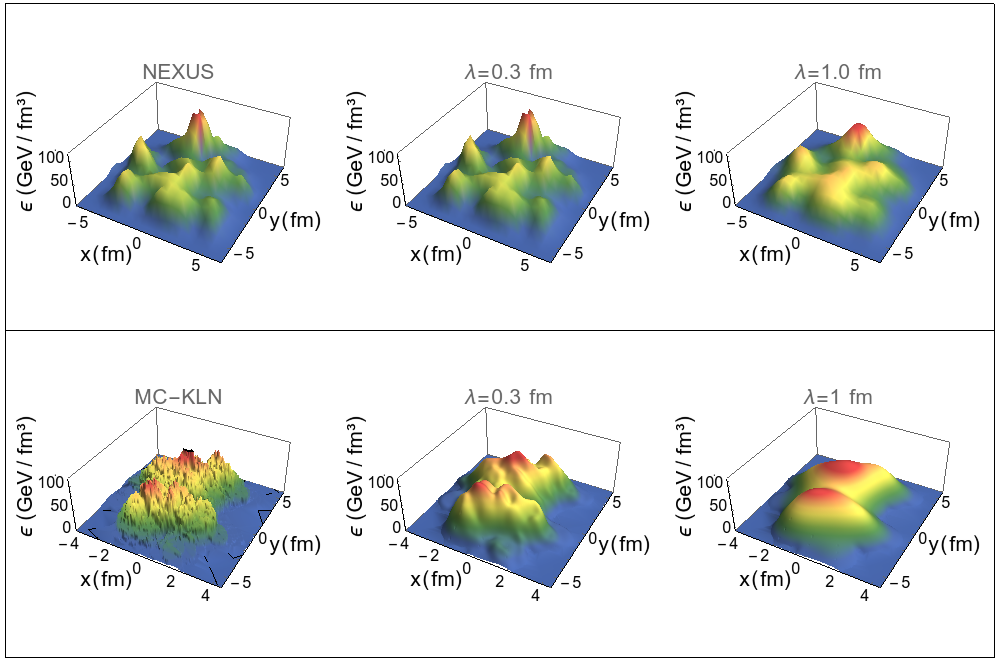}
\caption{
Initial energy density at midrapidity for a Pb-Pb collision at $\sqrt{s_{NN}}=2.76$ TeV without modification (left) and with a cubic spline filter using $\lambda$=0.3 and 1 fm for NEXUS initial conditions (top) and MCKLN initial conditions (bottom). }
  \label{fig:IC} 
\end{center}
\end{figure}

Small scale structure is systematically smoothed out using a cubic spline filter, W, where the transverse energy density, $\varepsilon$, at any point in space is determined through a weighted sum of energy density values at $\vec{r}_\alpha$ around it in the transverse plane:
 \begin{equation}
   \epsilon(\tau_0, \vec{r}; \lambda)=\sum_{\alpha=1}^{N} \epsilon(\tau_0,\vec{r_\alpha}) W\left(\frac{|\vec{r}-\vec{r_\alpha}|}{\lambda};\lambda\right)
 \end{equation}
A cubic spline ensures that its nearest neighbors contribute the most  (i.e. W peaks at  $\vec{r}-\vec{r_\alpha}=0$)  and that no contributions are considered beyond a distance of $2\lambda$.  Additionally, $\int W\left(\frac{|\vec{r}|}{\lambda};\lambda\right) d\vec{r}=1$
to ensure that the integral of $\epsilon(\tau_0, \vec{r}; \lambda)$ is not modified by changes in $\lambda$.
Increasing $\lambda$ consistently smooths out small scale structure within initial conditions \cite{Noronha-Hostler:2015coa}, as shown in Figure  \ref{fig:IC}.

In order to quantify the effect of the coarse-graining, we decompose the initial energy density  in a way that is ordered according to length scale  
\cite{Teaney:2010vd}. In this way, an eccentricity $\epsilon_n$ is basically the ratio of the lowest cumulant $W_{n,n}$
(which gives information on the largest-scale global structure)
by $\langle r^n \rangle$. We found that the $W_{n,n}$'s are unaffected by smoothing out $\lambda$, while higher-order cumulants more strongly depend on the smoothing parameter $\lambda$, with increasing sensitivity for larger $m$, as expected. However the smoothing process does have a small effect on $\langle r^n \rangle$ 
such that the magnitude of eccentricities decrease. Thus, any observables whose effect can be explained by this decrease in the eccentricities is not sensitive to the magnitude of small scale structure but rather on the large scale structure quantified by the magnitude of the eccentricities. Thus, if quantities that are rescaled by their eccentricities are affected beyond what is probable just from statistical uncertainty, this can be an indication that there is a dependence on the coarse graining of small scale structure.  In \cite{Gardim:2017ruc} we checked that both the eccentricities and their scaled event-by-event distributions are not significantly affected by the small scale $\lambda$.

\section{Results for observables}

Here we used the hydrodynamic models NeXSPheRIO \cite{Hama:2004rr} with NEXUS initial conditions \cite{Drescher:2000ha} and v-USPhydro  \cite{Noronha-Hostler:2014gga,Noronha-Hostler:2014dqa} coupled to 
  MC-KLN initial conditions \cite{Drescher:2006ca}.  
  
  It is well know that for $n=2,3$ that there is roughly linear scaling between $v_n \propto \varepsilon _n$ on an event-by-event basis, however, non-linear scaling can occur in peripheral collisions
  \cite{Niemi:2012aj,Noronha-Hostler:2015dbi}.  Thus, we compared the event-by-event distributions of flow harmonics and their dependence on $\lambda$ but we found no strong dependence across a variety of centralities.

One could argue that the $v_n$ distributions are only dependent on a single flow harmonic and mixed harmonic correlations may be more sensitive to small scale structure, as first measured in \cite{Zhou:2015slf,ALICE:2016kpq}. Here we used the same methodology including multiplicity weighing and centrality rebinning as discussed in \cite{Gardim:2016nrr}.  While there is not always linear scaling between symmetric cumulants and their eccentricities, we found an insensitivity to $\lambda$ for various normalized symmetric cumulants using both NeXus and MC-KLN initial conditions. Similarly, we found that event plane correlations (as first measured by ATLAS in \cite{Aad:2014fla}) also do not exhibit a sensitivity to $\lambda$.  This is, however, unsurprising since a connection can be derived between the symmetric cumulants and the event plane correlations \cite{Giacalone:2016afq}.

Because we did not find clear evidence of a  sensitivity to $\lambda=0.3-1$ fm in $p_T$ integrated observables of all charged particles, we instead turn to more differential quantities. Previous work compared the spectra of coarse grained UrQMD and MCKLN initial conditions in \cite{Petersen:2010zt,Noronha-Hostler:2015coa} and found little effect for coarse graining less than 1 fm.  Additionally in \cite{Petersen:2010zt,Noronha-Hostler:2015coa,RihanHaque:2012wp} the differential flow harmonics $v_n(p_T)$'s were studied that also did not depend strongly on coarse graining.  

An alternative is to instead look at the factorization breaking \cite{Gardim:2012im} that utilizes the information of the Fourier coefficients of the event-averaged pair correlations $V_{n\Delta}(p_1,p_2)$ defined as
\begin{equation}
  r_n(p_1,p_2)=\frac{ V_{n\Delta}(p_1,p_2)}{\sqrt{ V_{n\Delta}(p_1,p_1) V_{n\Delta}(p_2,p_2)}}.  
  \label{eq:ratio}
 \end{equation}
 The factorization breaking was measured by CMS \cite{Khachatryan:2015oea} and ALICE \cite{Acharya:2017ino}. In a previous paper \cite{Kozlov:2014fqa} it was shown that $r_n$ is sensitive to coarse graining, although other considerations such as shear viscosity in ultra central collisions \cite{Heinz:2013bua} and bulk viscosity and hadronic rescattering are discussed in \cite{McDonald:2016vlt}.

\begin{figure}[!ht]
\begin{center}
  \includegraphics[width=1\textwidth]{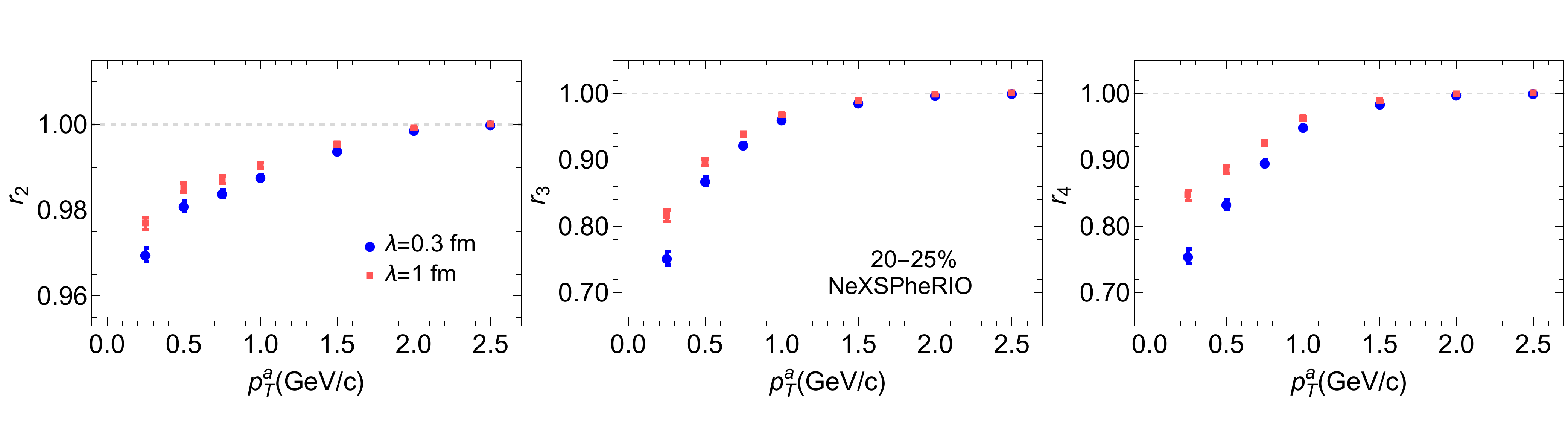} 
  \caption{Flow factorization ratio for NeXus with a range of $\lambda$'s in  20-25\% centrality and 
$2.5\,GeV<p_T^b<3.0\,GeV$.
 }
\label{fig:nexusrn}
\end{center}
\end{figure}

Our results for the factorization breaking from NEXUS initial conditions are  shown in Fig.\ \ref{fig:nexusrn}.  Note that the slight decrease in the eccentricities from $\lambda$ smoothing should approximately cancel out in Eq.~(\ref{eq:ratio}), so the effects is, indeed, from small scale structure. Thus, one can say with confidence that there is up to $\sim 15\%$ effect of small scale structure in the initial conditions on the factorization breaking.

As an additional check,   we carried out a
Principal Component Analysis  \cite{Bhalerao:2014mua}. Not surprisingly, we found that  the leading principal flow vector is almost independent of 
 the smoothing length whereas the  subleading principal flow vector does demonstrate some dependence on $\lambda$.

 \section{Conclusion}
 
 In order to answer the question of the influence of small scale structure within initial conditions on the final flow observables, we systematically filter a variety of initial conditions using a cubic spline such that global  large scale structures (e.g. eccentricities) are not significantly changed but small scale structure is filtered out. Using relativistic hydrodynamics, we find that a variety of integrated $v_n$ flow observables and $v_n$ distributions remain insensitive to small scale structure.  However, other more differential observables such as the factorization breaking and the subleading principle components do demonstrate a sensitivity to the small scale structure. Because the factorization breaking ratio is not strongly dependent on $\eta/s$, it is a strong candidate for distinguishing between initial conditions that exhibit different size hot spots.

\section{Acknowledgements}
We  thank  J.-Y.~Ollitrault for very helpful discussions on the PCA method. 
J.N.H acknowledges the use of the Maxwell Cluster and the advanced support from the Center of Advanced Computing and Data Systems at the University of Houston, and the support of the Alfred P. Sloan Foundation. 
F.G.~acknowledges  support  from
Funda\c{c}\~ao de Amparo \`a Pesquisa do Estado de S\~ao Paulo
(FAPESP  grants  2015/00011-8, 2015/50438-8,  2016/03274-2, 2018/00407-7), USP-COFECUB (grant Uc Ph 160-16 (2015/13) ), 
Conselho Nacional de Desenvolvimento Cient\'{\i}fico e Tecnol\'ogico (CNPq grant 310141/2016-8) and project INCT-FNA Proc.~No.~464898/2014-5.  F.G.G. was supported by Conselho Nacional de Desenvolvimento Cient\'{\i}fico  e  Tecnol\'ogico  (CNPq grant 312203/2015-2)  and FAPEMIG (grant APQ-02107-16).
 P.I.~thanks support from Coordena\c{c}\~ao de Aperfei\c{c}oamento de Pessoal de N\'{\i}vel Superior (CAPES).
M.L.~acknowledges support from FAPESP projects 2016/24029-6  and 2017/05685-2, and project INCT-FNA Proc.~No.~464898/2014-5.








\end{document}